\documentclass[11pt,prd,superscriptaddress,nofootinbib]{revtex4}
\usepackage[english]{babel}
\usepackage{graphicx}
\usepackage{mathtext}
\usepackage{indentfirst}
\usepackage{epsfig,amsmath,amsfonts}
\usepackage{color}

\newcommand{\beq}[1]{\begin{eqnarray}\label{#1}}
\newcommand\eeq {\end{eqnarray}}
\newcommand\bqa {\begin{eqnarray}}
\newcommand\eqa {\end{eqnarray}}

\newcommand{\bear}{\begin{array}}
\newcommand{\enar}{\end{array}}

\def\interior#1{\setbox1=\hbox{$#1$}\rlap{$#1$}\kern0.4\wd1\raise1.1\ht1%
\hbox{$\scriptstyle \circ$}}

\def\boxit#1#2{\setbox1=\hbox{\kern#1{#2}\kern#1}%
\dimen1=\ht1 \advance \dimen1 by #1 \dimen2=\dp1 \advance \dimen2 by #1
\setbox1=\hbox{\vrule height\dimen1 depth\dimen2\box1\vrule}%
\setbox1=\vbox{\hrule\box1\hrule}%
\advance \dimen1 by .4pt \ht1=\dimen1 \advance \dimen2 by .4pt \dp1=\dimen2
\box1\relax}
\def\endprf{\raise .5ex\hbox{\boxit{2pt}{\ }}}

\def\ifundefined#1{\expandafter\ifx\csname#1\endcsname\relax}

\def\beq{\begin{equation}}
\def\endq{\end{equation}}
\def\beqa{\begin{eqnarray}}
\def\endqa{\end{eqnarray}}

\begin{document}
\centerline{\bf \Large Corrections to the Aretakis type behaviour of the metric}
\centerline{\bf \Large due to an infalling particle}
\vspace{5mm}

\centerline{E. T. ${\rm Akhmedov}^{1, 2}$, P. A. ${\rm Anempodistov}^{3}$, I. D. ${\rm Ivanova}^{4}$}

\begin{center}
{\it $\phantom{1}^{1}$ Institutskii per, 9, Moscow Institute of Physics and Technology, 141700, Dolgoprudny, Russia}
\end{center}

\begin{center}
{\it $\phantom{1}^{2}$ B. Cheremushkinskaya, 25, Institute for Theoretical and Experimental Physics, 117218, Moscow, Russia}
\end{center}

\begin{center}
{\it $\phantom{1}^{3}$ National Research University MEPhI, 115409, Russia, Moscow, Kashirskoe shosse, 31}
\end{center}

\begin{center}
{\it $\phantom{1}^{4}$ National Research University HSE, 119048, Moscow, Usacheva str., 6.  Russian Federation}
\end{center}

\vspace{3mm}



\centerline{\bf Abstract}

We consider an extremal Reissner-Nordstr\"{o}m black hole perturbed by a neutral massive point particle, which falls in radially.
We study the linear metric perturbation in the vicinity of the black hole and find that the $l=0$ and $l=1$ spherical modes of the metric oscillate rather than decay.

\vspace{20mm}


{\bf 1.} Recently black hole physics in the vicinity of their horizons has received a lot of attention. This incudes the study of the BMS symmetries near the event horizon (see e.g. \cite{Hawking:2016msc}, \cite{Donnay:2015abr}, \cite{Donnay:2016ejv}, \cite{Akhmedov:2017ftb}), the study of non--extremal and extremal black holes in the vicinity of their horizons and critical surfaces, which was considered earlier in \cite{Price:1971fb}--\cite{Ferrell:1987gf} and recently continued in \cite{Hadar:2014dpa}--\cite{Camps:2017gxz} and the instability of extremal black holes under perturbations \cite{Durkee:2010ea}--\cite{Aretakis:2011hc}.

The study of the behaviour of metric perturbations in the vicinity of black holes is important for both understanding the mechanisms of the black hole creation and of their quantum radiation (see e.g \cite{Akhmedov:2016ati} for the general discussion). However, in the case of non extremal black holes the corresponding equations are complicated and are very hard to treat analytically. One is usually using computer simulations in such situations. Meanwhile a simple example which allows analytical treatment within this context is missing. At the same time usually via consideration of extremal black holes one is capable to make analytic observations \cite{Hadar:2014dpa}--\cite{Camps:2017gxz}. Of course one should bare in mind that such black holes are stable with respect to the quantum radiation, i.e. their consideration has a limiting applicability. The goal of the present paper is to consider metric perturbations in the vicinity of an extremal Reissner-Nordstr\"{o}m black hole caused by an infalling particle.

The Aretakis type of behaviour is that high enough derivative of a field (that could be a scalar, vector or tensor/metric field) grows as a power of time along the future horizon \cite{Durkee:2010ea}--\cite{Aretakis:2011hc}. Such a behaviour follows from a homogeneous linearized equation describing perturbations over the critical black hole. To see the behaviour one has to specify an appropriate boundary conditions. Unlike that case we consider perturbations of the metric due to infalling particle, i.e. we consider an inhomogeneous equation due to the particle's stress--energy tensor on its right hand side. In the present paper we make an observation that such an inhomogeneous equation has a particular retarded solution that does not decay with time, but oscillates. We see the oscillations already of the metric field itself without taking its derivatives.

{\bf 2.} To have an analytic headway we consider an extremal Reissner-Nordstr\"{o}m black hole perturbed by a neutral point particle of mass $m$, which falls in radially. Moreover, we look for a solution in the very vicinity of the black hole.

The background metric has the following form:

\begin{equation}
ds^{2}=-f(r)dv^{2}+2dvdr+r^{2}d\Omega^{2}, \quad  {\rm and} \quad d\Omega^2 = d\theta^2 + \sin^2\theta \, d\varphi^2,
\end{equation}
where in the extremal case $f(r)=1-\frac{2M}{r}+\frac{Q^{2}}{r^{2}}=\frac{(r-M)^{2}}{r^{2}}$ and two horizons coincide at $r_{\pm }=M $.
To be on the safe side and to avoid the degeneration of metric we first make the following change of coordinates $v \mapsto \frac{v}{\lambda }$ and $r \mapsto M+\lambda r$, i.e. consider the metric as follows:

\begin{equation}\label{lambdmetr}
ds^{2} = \bar{g}_{\mu\nu} \, dx^\mu \, dx^\nu = -\frac{r^{2}}{(M+\lambda r)^{2}}dv^{2}+2dvdr+( M+\lambda r)^{2}d\Omega^{2},
\end{equation}
and only then do we take the limit $\lambda \rightarrow 0$ \cite{Bardeen:1999px}, \cite{Kunduri:2007vf}, \cite{Kunduri:2013ana}. Note that now $r=0$ is the position of the horizon of the extremal black hole.

Namely below with the use of this metric we derive the linearized Einstein equations and then take the  $\lambda \to 0$ limit and solve the resulting equations. The right hand side of these equations is given by the stress--energy tensor of the free falling particle. For simplicity from the very beginning we derive the geodesics of the free falling particle in the limit $\lambda = 0$. In this limit the background metric takes the form:

 \begin{equation} \label{critBH}
ds^{2} = -\frac{r^{2}}{M^{2}}dv^{2}+2dvdr+M^{2}d\Omega^{2}.
\end{equation}
Ingoing radial geodesics in this metric are as follows:

\begin{equation}\label{7}
u^{0}=\frac{M}{r^{2}} \, \Bigl[EM-\sqrt{E^{2}M^{2}-r^{2}}\Bigr], \quad {\rm and} \quad
u^{1}=-\frac{\sqrt{E^{2}M^{2}-r^{2}}}{M},
\end{equation}
where $E$ is the energy of the particle per unit mass, i.e. it is a dimensionless quantity.
From $du^1/du^0 = dr/dv$ one can find the relation between $r$ and $v$ coordinates along the geodesic:

\begin{equation}
V(r)=C_{0}-\frac{M}{E r}\Bigl[EM-\sqrt{E^2 M^2-r^2}\Bigl],
\end{equation}
where $C_{0}$ corresponds to the moment of plunge.

Because this is geodesic the corresponding stress--energy tensor of the point like particle, which moves along it, obeys the necessary conservation conditions in the background metric. It has only three non--zero components:

\begin{eqnarray}
T_{00}=\frac{m E^2}{u^0 M^{2}sin \theta }\delta\left[r-R(v)\right] \, \delta (\theta -\theta _{0}) \, \delta (\phi -\phi_0), \nonumber \\
T_{11}=\frac{m u^{0}}{M^{2}sin \theta }\delta \left[r-R(v) \right] \, \delta (\theta -\theta _{0}) \, \delta (\phi -\phi_0), \nonumber \\
T_{01}=\frac{-E m}{M^{2}sin \theta }\delta \left[r-R(v)\right] \, \delta (\theta -\theta _{0}) \, \delta (\phi -\phi_0), \label{SET}
\end{eqnarray}
where $R(v)$ is the inverse function of $V(r)$ and $m$ is the mass of the particle.
The goal of the present paper is to find linear deformations of the metric (\ref{lambdmetr}) by the stress--energy tensor of the point particle in the vicinity of $r=0$.

Namely we consider linear perturbations, $h_{\mu\nu}$, of the background metric (\ref{lambdmetr}), $g_{\mu\nu} = \bar{g}_{\mu\nu} + h_{\mu\nu}$, by the stress--energy tensor (\ref{SET}). We derive the equation for $\gamma_{\mu\nu} = h_{\mu\nu} - \frac12 \, \bar{g}_{\mu\nu} \, h_\alpha^\alpha$ in the gauge $\bar{\nabla}^\mu \, \gamma_{\mu\nu} = 0$.
The only non--trivial equations are for the $\gamma_{ij}$, $i,j = \overline{0,1}$ components of the perturbations. They look complicated, but in the limit $\lambda \to 0$ simplify to the following form:

\begin{eqnarray}
\square \gamma_{00}-\frac{4r}{M^2} \partial_{r}\gamma_{00}+\frac{4r}{M^2 }\partial_{v} \gamma_{01} \approx -16\pi T_{00}, \nonumber \\
\square \gamma_{01}+\frac{2r}{M^2}\partial_{v}\gamma_{11} -\frac{2}{M^{2}}\gamma_{01} \approx -16\pi T_{01}, \nonumber \\
\square \gamma_{11}+\frac{4r}{M^2}\partial_{r}\gamma_{11} +\frac{4}{M^2}\gamma_{11} \approx -16\pi T_{11}. \label{maineq}
\end{eqnarray}
The other components of the linearized Einstein equations are trivially satisfied if we set the $\gamma_{\mu\nu}$, for $\mu,\nu \neq \overline{0,1}$ to zero.
Here

\begin{eqnarray}
\square=\frac{1}{M^2}\partial_{r}(r^2\partial_{r})+2\partial^{2}_{vr}+\frac{1}{M^2 \sin\theta}\partial_{\theta}(\sin\theta \partial_{\theta})+\frac{1}{M^2 \sin^2\theta}\partial^{2}_{\phi\phi} = \nonumber \\ =
\frac{1}{M^2}\partial_{r}(r^2\partial_{r})+2\partial^{2}_{vr}+\frac{1}{M^2}\triangle_{\theta \phi},
\end{eqnarray}
where $\triangle_{\theta \phi}$ is the Laplacian on the unit two--dimensional sphere. We are interested in the particular solution of the system of equations (\ref{maineq}) in the vicinity of the critical radius, i.e. as $r \to 0$. The boundary conditions are taken into account by the appropriate choice of the Green function of the differential operator on the right hand side of the equation under consideration. We use the retarded Green function.

{\bf 3.} As one can see all the derivatives with respect to the angular variables in eq.(\ref{maineq}) fold into the angular part of the Laplacian $\triangle_{\theta \phi}$. Such a simplification happens because we consider the {\it radial} infall of the massive particle. The situation gets way more complicated for a non--radial infall and will be considered elsewhere.

As a result it is convenient to expand $\gamma_{ij}$ and the components of the stress-energy tensor in the spherical harmonics $Y_{l,n}(\theta, \varphi)$.
Also one can Fourier transform $\gamma_{ij}$ and the stress--energy tensor with respect to the $v$ variable:

\begin{equation}
\gamma_{ij}(v,r,\theta, \varphi) = \frac{1}{\sqrt{2\pi}} \, \int_{-\infty}^{+\infty} d\omega \sum_{l,n} Y_{l,n}(\theta, \varphi) \, e^{-i\,\omega\, [v - V(r)]} \, \gamma_{ij,ln} (\omega, r), \quad i,j = \overline{0,1}.
\end{equation}
Here we have defined the Fourier components of $\gamma_{ij,ln}(v,r)$ as $\gamma_{ij,ln}(\omega, r) \, e^{i \, \omega \, V(r)}$.

Performing all these transformations in the system of equations (\ref{maineq}) and taking the limit $r \rightarrow 0$ we obtain:

\begin{eqnarray}\label{star}
\left[r^2 \frac{d^2}{dr^2} - 2i\omega M^2 \frac{d}{dr} + \Bigl(2 \omega^2 M^2 V'(0)-b_{ij}\Bigr)\right] \, \gamma_{ij;ln} (\omega, r) \approx \tilde{T}_{ij;ln}(0),
\end{eqnarray}
where

\begin{eqnarray}
b_{00}=l(l+1), \quad b_{01}=l(l+1)+2, \quad b_{11}=l(l+1)-4,
\end{eqnarray}
and

\begin{eqnarray}
\widetilde{T}_{00;ln}(0) = - 8 \, \sqrt{2\pi} \, m \, Y^{*}_{l,n}(\theta_{0},\phi_{0}), \quad
\widetilde{T}_{11;ln}(0) = - \frac{2 \, \sqrt{2\pi} \, m}{E^3}  \, Y^{*}_{l,n}(\theta_{0},\phi_{0}), \nonumber \\
{\rm and} \quad \widetilde{T}_{01;ln}(0) = \frac{4 \,\sqrt{2\pi} \, m}{E} \,  Y^{*}_{l,n}(\theta_{0},\phi_{0}),
\end{eqnarray}
where we have used $u^0(0) = 1/2E$ and $u^1(0)= - E$, as follows from (\ref{7}).

As mentioned above we are interested in the particular solution of the system of equations (\ref{star}). The exact solution of this system of equations is as follows:	

\begin{eqnarray}
  \gamma_{00;ln} (\omega, 0) = \frac{-\tilde{T}_{00;ln}(0)}{[\omega^2  M^2 E^{-2}+l(l+1)]}, \nonumber \\
  \gamma_{11;ln} (\omega, 0) = \frac{-\tilde{T}_{11;ln}(0)}{[\omega^2  M^2 E^{-2}+l(l+1)-4]}, \nonumber \\
  \gamma_{01;ln} (\omega, 0) = \frac{-\tilde{T}_{01;ln}(0)}{[\omega^2  M^2 E^{-2}+l(l+1)+2]}, \label{1313}
\end{eqnarray}
where we have used that $V'(0) = -1/2E^2$. 

Having calculated $\gamma_{ij;ln}(\omega, 0)$ we invert the Fourier transformation over $v$. Then for the component $ \gamma_{11;ln} (\omega, 0)$, one obtains:

\begin{multline}
  \gamma_{11;ln}(v,0)=\frac{1}{\sqrt{2\pi}}\int^{+\infty}_{-\infty}\frac{-\tilde{T}_{11;ln}(0)}{[\omega^2  M^2 E^{-2}+l(l+1)-4]}e^{i\omega [C_0-v]}d\omega = \\
	=\frac{-\tilde{T}_{11;ln}(0) E}{M}\sqrt{\frac{\pi}{2(l(l+1)-4)}} \, e^{-\frac{E}{M}\sqrt{l(l+1)-4}\left[v-C_0\right]} \, \vartheta\left[v-C_0\right],
\end{multline}
where $\vartheta\left[v-C_0\right]$ is the Heaviside theta--function and we have used that $V(0)= C_0$. We have kept here only the retarded solution. We see that after the plunge the perturbations of these components of the metric fade away with the time $v$.

However, such a form of $\gamma_{11;ln}(v,0)$ is valid for all $ l $, except $ l = 0,1 $. This is the usual story for the gravitational perturbations. In the latter case the integrand has two poles on the real axis. Taking this integral in the complex $\omega$--plane, we should deform the contour of integration in such a way that it corresponds to the retarded solution (i.e. it should go above the poles of the integrand). As a result we obtain the oscillating solution:

\begin{multline}
\gamma_{11;ln}(v,0)=\frac{\widetilde{T}_{11;ln}(0) E}{M} \, \sqrt{\frac{2\pi}{4-l(l+1)}} \, \sin\left[\frac{E}{M}\sqrt{4-l(l+1)}\left[v-C_0\right]\right] \, \vartheta\left[v-C_0\right], \quad l = 0,1.
\end{multline}
The remaining components of $\gamma_{ij;ln}(v, 0)$ are as follows:

\begin{eqnarray}
 \gamma_{00;ln}(v,0) = \frac{-\widetilde{T}_{00;ln}(0) \, E}{M} \, \sqrt{\frac{\pi}{2l(l+1)}} \, e^{-\frac{E}{M}\sqrt{l(l+1)}\left[v-C_0\right]} \, \vartheta\left[v-C_0\right], \nonumber \\
\gamma_{01;ln}(v,0)=\frac{-\widetilde{T}_{01;ln}(0) E}{M} \, \sqrt{\frac{\pi}{2[l(l+1)+2]}} \, e^{-\frac{E}{M}\sqrt{l(l+1)+2} \,\left[v-C_0\right]} \, \vartheta\left[v-C_0\right].
\end{eqnarray}
All other components of $\gamma_{\mu\nu}$ are vanishing.

{\bf 4.} Thus, for all values of $l$ all components of $\gamma_{\mu\nu;ln}(v, 0)$ either decay in time $v$ or vanish from the very beginning. The exceptions are the component $\gamma_{11;ln}(v, 0)$ for $l = 0,1$, which just oscillate after the moment of plunge and do not decay. The presence of oscillations in $\gamma_{11;ln}(v, 0)$ means their presence in $h_{00;ln}(v, 0)$, $h_{01;ln}(v, 0)$ and $h_{11;ln}(v, 0)$ for $l=0,1$. This is a very unusual behaviour, because according to the early study of non--extremal black hole perturbations \cite{Price:1971fb}, \cite{Price:1972pw} one would expect that linear perturbations should probably fade away with time. However, in the case of extremal black holes there can be some unexpected type of behaviour \cite{Aretakis:2011ha}, \cite{Aretakis:2011hc}.

Let us discuss what are the possible physical consequences of the fact that linear perturbations over the critical black hole do not decay with time. First, to better understand the meaning of the obtained result one has to consider the full equation describing linear perturbations of the metric (\ref{lambdmetr}):

\begin{equation}\label{entireeq}
\left[\partial_{r} r^2\partial_{r} - 2i\omega(M+\lambda r)^2\partial_{r} +\frac{4a_{ij} r M}{(M+\lambda r)} \partial_{r}-V_{ij}(r,\omega)\right] \gamma_{ij;ln}(\omega, r) = (M+\lambda r)^2 \, F_{ij;ln}(\omega,r),
\end{equation}
where

\begin{equation}
V_{ij}(r,\omega)=2i\omega \lambda(M+\lambda r) + l(l+1) + 2 b_{ij} + \frac{2 c_{ij}}{(M+\lambda r)}+\frac{2M d_{ij}}{(M+\lambda r)^2},
\end{equation}
and
$$
a_{00}=-1, \quad a_{01}=0, \quad a_{11}=1,
$$
$$
b_{00}=0, \quad b_{01}=1, \quad b_{11}=2,
$$
$$
c_{00}=1, \quad c_{01}=-2, \quad c_{11}=-1,
$$
$$
d_{00}=-1, \quad d_{01}=2, \quad d_{11}=-1,
$$
Also here $F_{ij;lm}(\omega,r)$ is the source due to the stress--energy tensor of the free falling particle. 

This equation is valid for any values of $r$ and for the entire critical black hole metric (\ref{lambdmetr}). To approximately solve such an equation one can apply the approach used e.g. in \cite{Casals:2016mel}. But that demands a separate study. At this moment, however, it is straightforward to show that corrections to (\ref{1313}) decay with $r \to 0$. That can be seen via the Taylor expansion in powers of $r$ of the corresponding solution of (\ref{entireeq}).

Second, of course to completely understand the physical consequences of the observed phenomenon one has to study the perturbation in the full non--linear theory. Probably this sort of a problem is beyond analytical methods. However, even now on general grounds one can try to guess the answer. In fact, observing that only $l=0$ and $l=1$ perturbations do not fade away one can expect that as the consequence of the fallen particle the black hole acquires extra mass or even angular momentum, i.e. becomes non--critical. In the case of the latter type of black holes all linear perturbations already fade away.

We would like to thank M.Godazgar for very valuable discussions. The work of ETA was supported by the state grant Goszadanie 3.9904.2017/BCh, by the Foundation for the Advancement of Theoretical Physics and Mathematics “BASIS” and by the grant RFBR 18-01-00460.

\end{document}